\documentclass[twocolumn]{aastex631}

\DeclareUnicodeCharacter{2212}{-}
\usepackage{multirow}
\usepackage{comment}
\usepackage[normalem]{ulem}
\usepackage{amsmath}

\shorttitle{Main-Sequence stars in NGC 188}
\shortauthors{Sun et al.}

\graphicspath{{./}{figures/}}
	
\begin{document}
		
\title{Lithium Abundances of Main-Sequence Stars in the Old Open Cluster NGC 188: Probes of Stellar Evolution Beyond the Solar Age\footnote{This is paper 94 of The WIYN Open Cluster Study.}}
		
\correspondingauthor{Qinghui Sun}
\email{qinghuisun@sjtu.edu.cn}
		
\author[0000-0003-3281-6461]{Qinghui Sun}
\affiliation{Tsung-Dao Lee Institute, Shanghai Jiao Tong University, Shanghai, 200240, China}

\author[0000-0002-3854-050X]{Constantine P. Deliyannis}
\affiliation{Department of Astronomy, Indiana University, Bloomington, IN 47405, USA}

\author[0000-0001-5436-5206]{Bruce A. Twarog}
\author[0000-0001-8841-3579]{Barbara J. Anthony-Twarog}
\affiliation{Department of Physics and Astronomy, University of Kansas, Lawrence, KS 660045, USA}

\begin{abstract}

We present lithium abundances for 119 main-sequence stars in the old open cluster NGC 188 (age = 6.3 Gyr), using high-resolution, high signal-to-noise ratio spectra from WIYN/Hydra. We observe the stars over multiple nights and measure radial velocities for each night, which we combine with Gaia proper motions to identify multiplicity and cluster membership. We identify 95 single members, 14 binary members, 9 members with uncertain multiplicity, and 1 single likely member. We determine effective temperatures using empirical color–temperature relations, surface gravities from isochrones, and microturbulence from empirical relations for main-sequence stars. Our sample includes G dwarfs with temperatures between 6000 and 5300 K, which expands significantly on earlier observations. We find that lithium abundances in NGC 188 are lower than predictions from standard stellar evolution theory. As expected, stars in NGC 188 are more lithium-depleted than those in the younger Hyades and Praesepe clusters (650 Myr). However, their abundances are higher, or at least comparable, to those in the slightly younger cluster M67 (4 Gyr), challenging the idea that older stars have lower lithium than younger ones. Lithium depletion may depend on factors beyond age and mass, such as metallicity.
			
\end{abstract}
		
\section{Introduction}

Lithium (Li) is easily destroyed at relatively low temperatures ($\sim$2.5 MK; \citealt{1990AJ.....99..595S}) in stellar interiors, making its surface abundance a sensitive diagnostic of internal stellar processes. According to standard stellar evolution theory (SSET; \citealt{1967ARA&A...5..571I, 1990ApJS...73...21D, 2017AJ....153..128C}), Li is depleted at the base of the surface convection zone and only during the pre-main sequence for G dwarfs. SSET predicts progressively greater Li depletion in cooler dwarfs, which have deeper convective envelopes where Li is depleted at its hotter base. However, SSET does not account for additional processes such as rotation, magnetic fields, mass loss, or diffusion, all of which may further influence surface Li abundances (A(Li)).

Observations, however, reveal that many dwarfs show A(Li) patterns that deviate significantly from SSET predictions. Notable examples include the F-dwarf Li-Dip, a prominent and highly non-standard feature (e.g., \citealt{2025arXiv250704266S, 2025arXiv250808671S}), and the over-depletion of Li in older G-type dwarfs \citep[e.g.,][]{2023ApJ...952...71S}. The Sun itself displays a surface A(Li) approximately 50 times lower than that predicted by SSET \citep{1997AJ....113.1871K}. These discrepancies suggest that additional physical processes beyond those included in SSET should contribute to Li depletion. Several non-standard mechanisms have been proposed to account for this, including mass loss \citep{1990ApJ...359L..55S}, microscopic diffusion \citep{1986ApJ...302..650M, 1993ApJ...416..312R}, and rotationally induced mixing \citep{1989ApJ...338..424P, 1990ApJS...74..501P}.

Open clusters, which are co-eval groups of stars with a common origin and similar chemical compositions, serve as valuable laboratories for studying physical processes in stellar interiors \citep{2025Natur.640..338R}, the time evolution of stars \citep{2019ARA&A..57..227K}, and broader topics such as potential connections to exoplanet formation and occurrence \citep{2022RAA....22g5008S, 2023ApJ...952...68S}. In particular, A(Li) patterns observed in cluster stars, which span a wide range of effective temperatures ($T_{\rm eff}$) and rotation rates but share the same age and metallicity, provide strong observational constraints on the internal physics of stars. Different non-standard mixing mechanisms produce distinct signatures in A(Li), allowing these processes to be examined in a systematic manner using open cluster populations.

Among the proposed mechanisms, rotational mixing has received strong observational support from the evolution of the A(Li)–$T_{\rm eff}$ pattern in G dwarfs in open clusters. In young clusters such as the Pleiades ($\sim$100 Myr; \citealt{2015MNRAS.449.4131S}) and M35 ($\sim$150 Myr; \citealt{2021MNRAS.500.1158J}), G dwarfs show decreasing A(Li) with decreasing $T_{\rm eff}$, broadly consistent with predictions from standard stellar evolution theory (SSET). However, for fast rotators in these clusters, the observed A(Li) are higher than predicted by SSET. As clusters age, A(Li) continues to decline in cooler G dwarfs, resulting in both lower A(Li) and a steeper A(Li)–$T_{\rm eff}$ trend. This behavior is evident in intermediate-age clusters such as the Hyades and Praesepe ($\sim$650 Myr; \citealt{2017AJ....153..128C}) and in the older cluster M67 ($\sim$4 Gyr; \citealt{2000ApJ...544..944S, 2012MSAIS..22...97P}). The progressive depletion of A(Li) and the steepening of the A(Li)–$T_{\rm eff}$ trend with age support that rotational mixing plays a key role in shaping surface A(Li) in G dwarfs, as suggested by previous studies \citep{2017AJ....153..128C, 2023ApJ...952...71S}.

In this paper, we present a detailed analysis of A(Li) in main-sequence G dwarfs in NGC 188, one of the oldest open clusters in the Milky Way (6.3 Gyr; \citealt{2022MNRAS.513.5387S}), based on new spectroscopic observations obtained with WIYN/Hydra. As the oldest open cluster studied for the purpose of examining the A(Li)–$T_{\rm eff}$ relation since M67, NGC 188 provides a valuable opportunity to investigate how the A(Li)–$T_{\rm eff}$ morphology evolves beyond the solar age. To date, only two studies have reported A(Li) for dwarfs in this cluster: one analyzed seven main-sequence stars \citep{1988ApJ...334..734H}, and the other examined eleven G-type stars \citep{2003A&A...399..133R}, with no new results published in the past two decades. The present study significantly expands the sample size, reporting A(Li) for 119 G dwarfs in NGC 188.
		
\section{Observations and Data Selection}

Spectroscopic data for NGC 188 candidate members, spanning from low-mass main-sequence stars to those on the red giant branch, were collected using the Hydra multi-object spectrograph on the WIYN 3.5-meter telescope. These observations were carried out across multiple runs, including those in November 1995, April 1996, June 1997, February and March 2001, April 2002, and December 2017, using various fiber configurations. Most of the data were obtained with the 316 l/mm echelle grating, which yielded a spectral resolution of about R $\sim$ 13,000 with the blue fiber set and R $\sim$ 17,000 with the red, covering a wavelength window of roughly 400 \AA\ centered near 6650 \AA. For the runs in March 2001 and April 2002, we used the 31.6 l/mm KPNO coude grating instead, which offered improved efficiency and slightly higher resolution (R $\sim$ 19,000) over the narrow range of 6700–6730 \AA. While this setup is well suited for observing the Li I 6707.78 \AA\ feature, its limited number of Fe I lines makes it unsuitable for reliable metallicity measurements, which were instead derived from spectra taken with the 316 l/mm grating (\citealt{2022MNRAS.513.5387S}).

\begin{figure}
	\centering
	\includegraphics[width=0.45\textwidth]{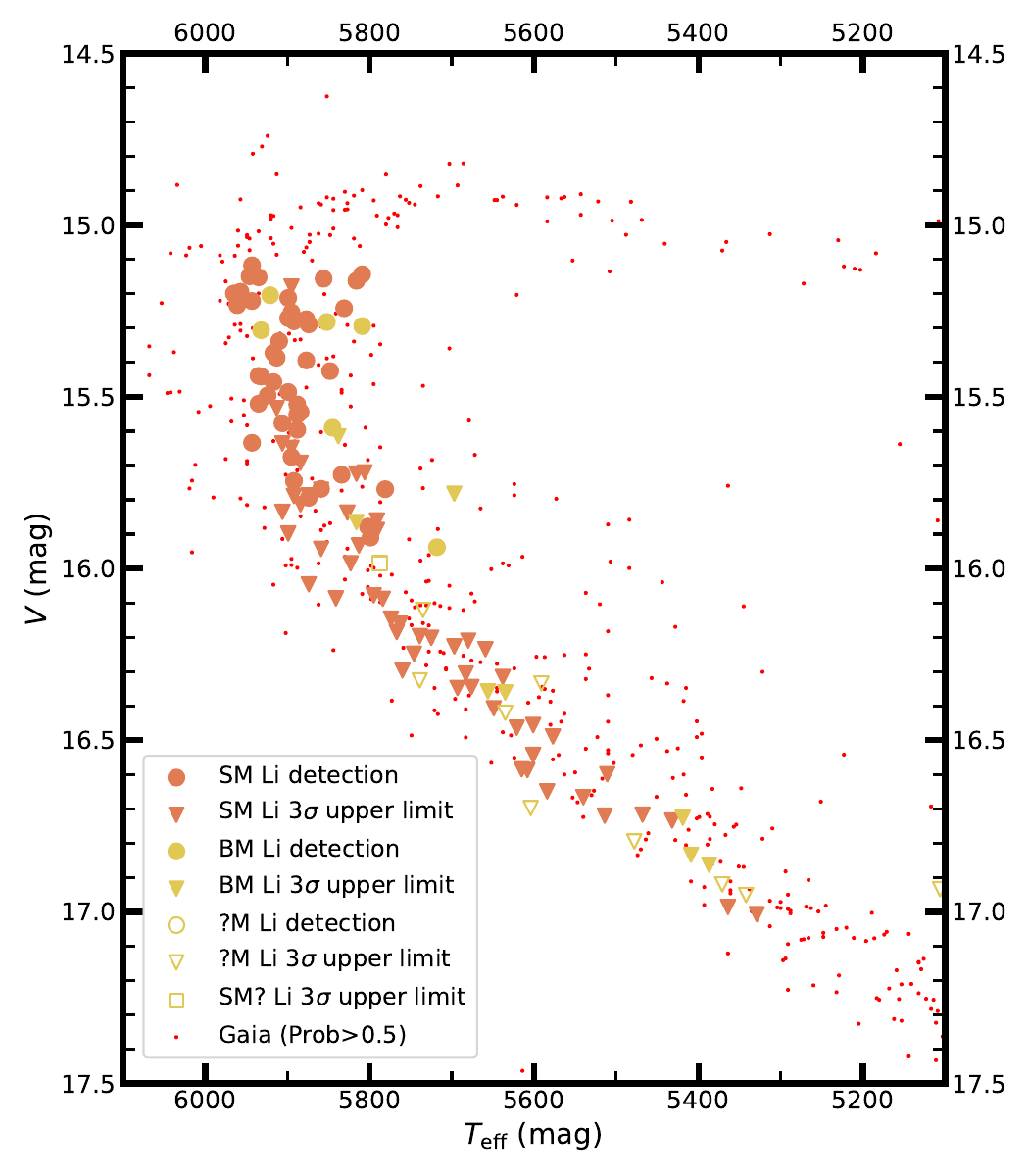}
	\caption{$V$ magnitude as a function of $T_{\rm eff}$ for main-sequence stars in NGC 188. The single members (SM), binary members (BM), members with uncertain multiplicity (?M), and single likely members (SM?) are shown in different symbols, as denoted in the legend. Stars with cluster membership probability greater than 0.5 from \citet{2018A&A...618A..93C} are shown in small red dots.}
	\label{fig1}
\end{figure}

Previous work by \citet{2022MNRAS.513.5387S} focused on brighter, evolved stars ($V<14.16$ mag), discussing possible Li enrichment in some red giants. More recently, \citet{2025arXiv250704266S} analyzed stars near the turnoff and on the subgiant branch ($14.16 \leq V \leq 15.4$ mag) to investigate the Li-Dip. In this study, we restrict the sample to stars fainter than $V=15.4$ mag, thereby including only MS stars in NGC 188. The targets were first selected from the color–magnitude diagram by requiring photometric membership, and then cross-matched with the proper motion study of \citet{2003AJ....126.2922P}, retaining only stars with membership probabilities greater than 70\%. The final sample includes 119 stars, which are shown in the $T_{\rm eff}$–magnitude diagram in Figure~\ref{fig1}; the corresponding stellar parameters are listed in Table~\ref{table1}. We plot $T_{\rm eff}$ instead of $B-V$, and details of the $T_{\rm eff}$ determination are given in Section~\ref{atmosphere}.

In Table~\ref{table1}, we show both the WIYN Open Cluster Study (WOCS) ID and the ID from \citet{2003AJ....126.2922P}. The WOCS nomenclature scheme was introduced in the study of NGC 6819 \cite{2009AJ....138..159H}. For NGC 188, we use the photometry data from Gaia DR3 (\citealt{2023AA...674A...1G}), converting $G$ to $V$ magnitude using the relationship specified in the Gaia DR3 documentation (``5.5.1 Relationships with other photometric systems"). We adopt the cluster center at RA = 11.83111 deg, DEC = 85.24118 deg (\citealt{2022MNRAS.513.5387S}). In the WOCS numbering system, which orders stars by increasing $V$ magnitude within concentric annuli (30" wide) centered on the cluster, each star's ID combines its brightness rank within an annulus (one-three digits) so that the ID itself gives some idea of the star’s brightness and location and the annulus number (also three digits); for example, star 1003 is the brightest in the third annulus.

\begin{deluxetable*}{cccccccccccccccccc}
	\label{table1}
	\tablecaption{Stellar parameters of NGC 188 main-sequence stars}
	\tabletypesize{\scriptsize}
	\setlength{\tabcolsep}{3pt}
	\decimalcolnumbers
	\renewcommand{\arraystretch}{1.0}
	\tablehead{
		\colhead{WOCS ID$^a$} & \colhead{ID$^b$} & \colhead{RA$^c$} & \colhead{DEC$^c$} & \colhead{$V^d$} & \colhead{$B-V^d$} & \colhead{$\sigma_{B-V}^d$} & \colhead{$T_{\rm eff}^e$}	& \colhead{$\sigma_{T_{\rm eff}}^e$} & \colhead{log g$^e$} & \colhead{$V_{\rm t}^e$} & \colhead{$V_{RAD}^f$} & \colhead{$\sigma_{V_{RAD}}^f$} & \colhead{mm$^g$} & \colhead{A(Li)$^h$} & \colhead{$\sigma_{A(Li)}^h$} & \colhead{S/N$^h$} & \colhead{RUWE$^i$} \\
		\colhead{} & \colhead{deg} & \colhead{deg} & \colhead{mag} & \colhead{mag} & \colhead{mag} & \colhead{K}	& \colhead{K} & \colhead{} & \colhead{km s$^{-1}$} & \colhead{km s$^{-1}$} & \colhead{km s$^{-1}$} & \colhead{} & \colhead{dex} & \colhead{dex} & \colhead{} & \colhead{}
	} 
	\startdata
	\hline
	6003  & 5083 & 11.83299  & 85.22215 & 15.117  & 0.679 & 0.007 & 5943 &  26 & 4.08 & 1.74 & -43.98 & 0.90 & sm & 2.50    & 0.09 & 65  & 0.963 \\
	12032  & 6064 & 14.08171  & 85.42691 & 15.143  & 0.716 & 0.013 & 5809 &  46 & 4.09 & 1.63 & -42.08 & 0.65 & sm & 2.20    & 0.07 & 121 & 0.987 \\
	10006  & 5338 & 12.27250  & 85.27061 & 15.149  & 0.678 & 0.006 & 5946 &  22 & 4.10 & 1.73 & -42.77 & 0.73 & sm & 2.40    & 0.04 & 133 & 0.971 \\
	...  & ... & ...  & ... & ...  & ... & ... & ... &  ... & ... & ... & ... & ... & ... & ... & ... & ...  & ... \\
	\hline
	\enddata
	\tablecomments{a. WIYN Open Cluster Study (WOCS) ID. \\
		b. ID from \citet{2003AJ....126.2922P}. \\
		c. RA and DEC (J2000) from Gaia DR3 (\citealt{2023AA...674A...1G}). \\
		d. The average $V$ and $B − V$. The dash in $\sigma_{B-V}$ denotes that the star has only one $B-V$ magnitude measurement. \\
		e. Stellar atmospheric parameters including $T_{\rm eff}$, log g, and microturbulence ($V_{\rm t}$). \\
		f. Radial velocity ($V_{\rm RAD}$) and its error ($\sigma_{V_{\rm RAD}}$), are obtained from the final combined spectrum by using the {\it fxcor} task in IRAF. The combined $V_{\rm RAD}$ values are determined solely for single stars. The rotational velocity ($V_{\rm ROT}$) has an upper limit of 20 km s$^{-1}$, and is not explicitly shown here for individuals. \\
		g. Final multiplicity and membership (mm). \\
		h. A(Li) and its error. When ``$<$" is used, it indicates a 3$\sigma$ upper limit for the abundance, with no separate error provided for upper limits as they inherently include error. The final errors are co-added in quadrature from those propagated from the stellar atmosphere and the 1$\sigma$ equivalent width measurement. \\
		i. Gaia DR3 Re-normalized Unit Weight Error (RUWE) values. RUWE does not always agree with our multiplicity classifications based on radial velocities.
		\\ (This table is available in its entirety online)}
\end{deluxetable*}

\section{Radial velocities, membership, and rotational velocities}

Radial velocities ($V_{\rm RAD}$s) were measured for each night using the IRAF\footnote{IRAF is distributed by the National Optical Astronomy Observatories, operated by the Association of Universities for Research in Astronomy Inc., under a cooperative agreement with the National Science Foundation.} task \textit{fxcor}, following methods described in detail in \citet{2020AJ....159..220S, 2022MNRAS.513.5387S}. We combined these measurements with Gaia DR3 \citep{2023AA...674A...1G} proper motions and parallaxes to evaluate both cluster membership and stellar multiplicity. 

In brief, a star was classified as a cluster member if its parallax and proper motions in RA and Dec lay within 2$\sigma$ of the cluster mean values. We adopt the mean astrometric parameters from \citet{2022MNRAS.513.5387S}: parallax $=0.509 \pm 0.043$ mas, proper motion in RA $=-2.313 \pm 0.139$ mas yr$^{-1}$, and proper motion in Dec $=0.952 \pm 0.140$ mas yr$^{-1}$. The quoted uncertainties are standard deviations derived from Gaussian fits to the respective distributions (see Figure 3 of \citealt{2022MNRAS.513.5387S}), and twice these values (2$\sigma$) were used as our membership thresholds. Radial velocities served as an additional criterion for single stars: a star was considered a member if its $V_{\rm RAD}$ lay within 2$\sigma$ of the cluster mean of –42.89 km s$^{-1}$, where the 2$\sigma$ value corresponds to twice the individual measurement uncertainties (shown in Table \ref{table2}). For binaries, $V_{\rm RAD}$ variations were detected across multiple nights, but their parallaxes and proper motions remained consistent with cluster membership. In total, we treated parallax, proper motion in RA, proper motion in Dec, and $V_{\rm RAD}$ as four independent criteria. Single stars satisfying all four were classified as members, those meeting three were classified as likely members, and those meeting fewer were considered non-members or uncertain. Stars meeting the three astrometric criteria but showing multi-epoch $V_{\rm RAD}$ variations were classified as binary members.

We also compared our results with \citet{2008AJ....135.2264G}, who conducted long-term $V_{\rm RAD}$ monitoring in NGC 188, and found good agreement between our assignment and theirs. Based on these criteria, we identified 95 stars as single members (SM), 14 as binary members (BM), 9 as members with uncertain multiplicity (?M), and 1 as a single likely member (SM?). The designation scheme follows those used in \citet{2020AJ....159..220S}. Stellar parameters are listed in Table~\ref{table1}, and per-night $V_{\rm RAD}$s are provided in Table~\ref{table2}. Projected rotational velocities ({\it v} sin {\it i}) were also estimated using \textit{fxcor} by correlating with template spectra. All NGC 188 main-sequence stars in our sample show {\it v} sin {\it i} below the detection limit of 20 km s$^{-1}$ for our instrument and setup, as suggested in earlier work \citep[e.g.,][]{2022MNRAS.513.5387S, 2023ApJ...952...71S}. These {\it v} sin {\it i} values are therefore not reported in Table~\ref{table1}.

\begin{deluxetable*}{cccccccccccccccccccc}
	\label{table2}
	\tablecaption{Radial velocity from each individual night}
	\tabletypesize{\scriptsize}
	\setlength{\tabcolsep}{3pt}
	\decimalcolnumbers
	\tablehead{
		\colhead{ID$^a$} & \colhead{$V_{\rm RAD}^b$}	& \colhead{$\sigma^b$} & \colhead{con$^b$} & \colhead{$V_{\rm RAD}$}	& \colhead{$\sigma$} & \colhead{con} &  \colhead{$V_{\rm RAD}$}	& \colhead{$\sigma$} & \colhead{con} & \colhead{$V_{\rm RAD}$}	& \colhead{$\sigma$} & \colhead{con} & \colhead{$V_{\rm RAD}$}	& \colhead{$\sigma$} & \colhead{con} & \colhead{$V_{\rm RAD}$}	& \colhead{$\sigma$} & \colhead{con} &	\colhead{...} \\
		\colhead{} & \colhead{n1}	& \colhead{n1} & \colhead{n1} & \colhead{n2}	& \colhead{n2} & \colhead{n2} &  \colhead{n3}	& \colhead{n3} & \colhead{n3} & \colhead{n4}	& \colhead{n4} & \colhead{n4} & \colhead{n5}	& \colhead{n5} & \colhead{n5} & \colhead{n6}	& \colhead{n6} & \colhead{n6} &	\colhead{...}
	} 
	\startdata
	\hline
	5083	&	-41.33	&	1.77	&	1	&	-45.1	&	2.92	&	2	&	-44.59	&	1.18	&	4	&	-44.07	&	1.33	&	5	&	-44.69	&	1.53	&	6	&	--	&	--	&	--	&	...	\\
	6064	&	-42.38	&	0.83	&	18	&	-41.92	&	1.24	&	19	&	--	&	--	&	--	&	--	&	--	&	--	&	--	&	--	&	--	&	--	&	--	&	--	&	...	\\
	5338	&	-42.88	&	0.93	&	18	&	-42.33	&	1.84	&	19	&	--	&	--	&	--	&	--	&	--	&	--	&	--	&	--	&	--	&	--	&	--	&	--	&	...	\\
	...	&	...	&	...	&	...	&	...	&	...	&	...	&	...	& ...	&	...	&	...	& ...	&	...	&	...	&	...	&	...	&	...	&	...	&	...	 &	...	\\
	\hline
	\enddata
	\tablecomments{a. ID from \citet{2003AJ....126.2922P}. \\
		b. Radial velocity ($V_{\rm RAD}$), error in radial velocity ($\sigma$), and the corresponding WIYN configuration (con), , which are: 1 –1995 November 13, 2 – 1995 November 14, 3 –  1995 November 15, 4 - 1996 April 27, 5 - 1996 April 28, 6 - 1996 April 29, 7 - 1997 June 7, 8 - 1997 June 8, 9 - 2001 February 21, 10 - 2001 March 20, 11 - 2001 March 21, 12 - 2001 March 22, 13 - 2002 April 2, 14 - 2002 April 4, 15 - 2002 April 5, 16 - 2002 April 6, 17 - 2002 April 7, 18 – 2017 December 27, and 19 – 2017 December 28. \\
		\\ (This table is available in its entirety online)}
\end{deluxetable*}

\section{Stellar Atmospheres} \label{atmosphere}

We followed the procedures detailed in \citet{2022MNRAS.513.5387S} to derive the average $V$ and $B - V$ magnitudes. Photometry from several previous studies of NGC 188 \citep{2003AJ....126.2922P, 1999AJ....118.2894S, 2000AAS...196.4208H} was cross-calibrated to a common system and then averaged. For each star, we computed all ten possible color indices from $UBVRI$, then converted them into equivalent $B - V$ values using empirical polynomial relations calibrated on cluster members. The final $B - V$ value is the mean of these ten estimates, which reduces random errors to $\sigma(B-V) \sim 0.01$–$0.02$ mag. The standard deviation, $\sigma_{B-V}$, reflects the spread among the ten computed values and is listed in Table~\ref{table1}. A dash in this column indicates that only one $B-V$ value was available, so no standard deviation could be calculated.

Stellar atmospheric parameters, including effective temperature ($T_{\rm eff}$), surface gravity (log g), and microturbulence ($V_{\rm t}$), were determined following the same approach described in \citet{2020AJ....159..220S}. We adopted a metallicity of [Fe/H] = +0.064 $\pm$ 0.018 dex and used a 6.3 Gyr $Y^2$ isochrone \citep{2004ApJS..155..667D} with $E(B-V) = 0.09$ mag (\citealt{1999AJ....118.2894S}). $T_{\rm eff}$ were derived from empirical de-reddened $B-V$–$T_{\rm eff}$ relations from \citet{2023ApJ...952...71S} for MS stars. Uncertainties in $T_{\rm eff}$ were propagated from the $\sigma(B-V)$ values. Surface gravity was estimated by comparing the stars' positions on the CMD to the $Y^2$ isochrone \citep{2001ApJS..136..417Y} that best matched each star’s $T_{\rm eff}$. The associated log g errors were propagated from the $T_{\rm eff}$ uncertainties, and microturbulence uncertainties were propagated from both $T_{\rm eff}$ and log g. The resulting stellar parameters are provided in Table~\ref{table1}.

\section{Lithium Abundances}

Our abundance analysis follows the same procedures described in \citet{2022MNRAS.513.5387S, 2023ApJ...952...71S, 2025ApJ...978..107S}. Briefly, we generated synthetic spectra around the Li I 6707.8 \AA\ feature for all 119 stars using MOOG and derived lithium abundances expressed as A(Li) = 12 + log(N$_{\rm Li}$/N$_{\rm H}$). To distinguish Li detections from non-detections, we applied the $3\sigma$ criterion outlined by \citet{1993ApJ...414..740D}. For detections, A(Li) values were determined through spectral synthesis, while $3\sigma$ upper limits for non-detections were estimated using local signal-to-noise ratios (S/N) and line widths. We adopted the updated line list from \citet{2022MNRAS.513.5387S} and did not include $^6$Li, which is expected to be significantly depleted in these stars. Final A(Li) and upper limits (denoted with ``$<$'') are listed in Table~\ref{table1}. For detections, A(Li) uncertainties include contributions from both stellar atmosphere parameter errors and 1$\sigma$ equivalent width uncertainties, added in quadrature, following the approach used in \citet{2024AJ....167..167S, 2025ApJ...980..179S}. For stars with upper limits, we do not report separate uncertainties, as the values already include measurement errors.

\begin{figure}
	\centering
	\includegraphics[width=0.45\textwidth]{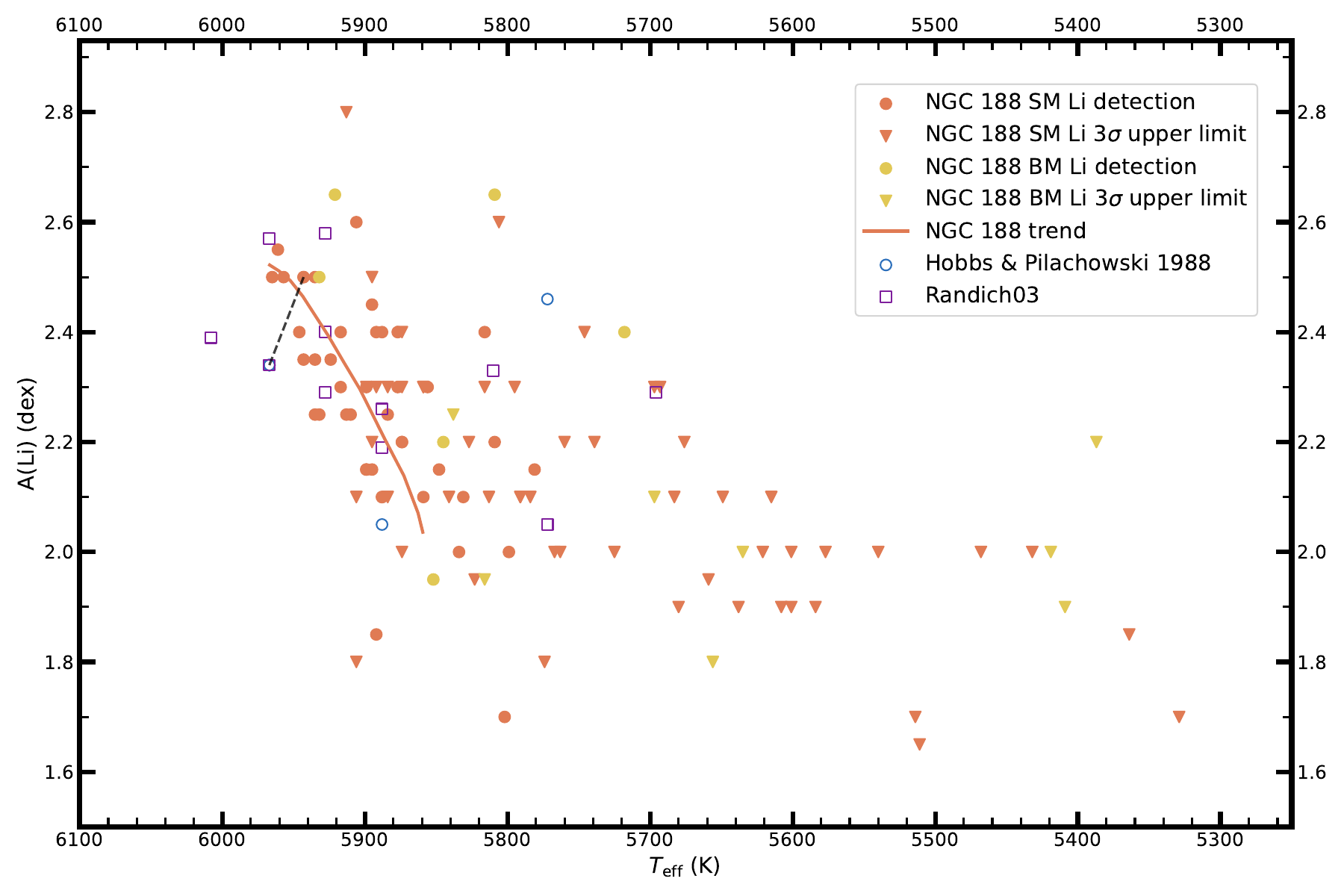}
	\caption{A(Li) as a function of $T_{\rm eff}$ for main-sequence stars in NGC 188. The same symbols are used for NGC 188 as in Figure~\ref{fig1}, while only SM and BM are shown. For comparison, we show A(Li) in NGC 188 from the literature (\citealt{1988ApJ...334..734H, 2003A&A...399..133R}), with the one common star linked by a gray line. The circles are A(Li) detections, while the downward triangles are A(Li) 3$\sigma$ upper limits.}
	\label{fig2}
\end{figure}

\begin{figure}
	\centering
	\includegraphics[width=0.45\textwidth]{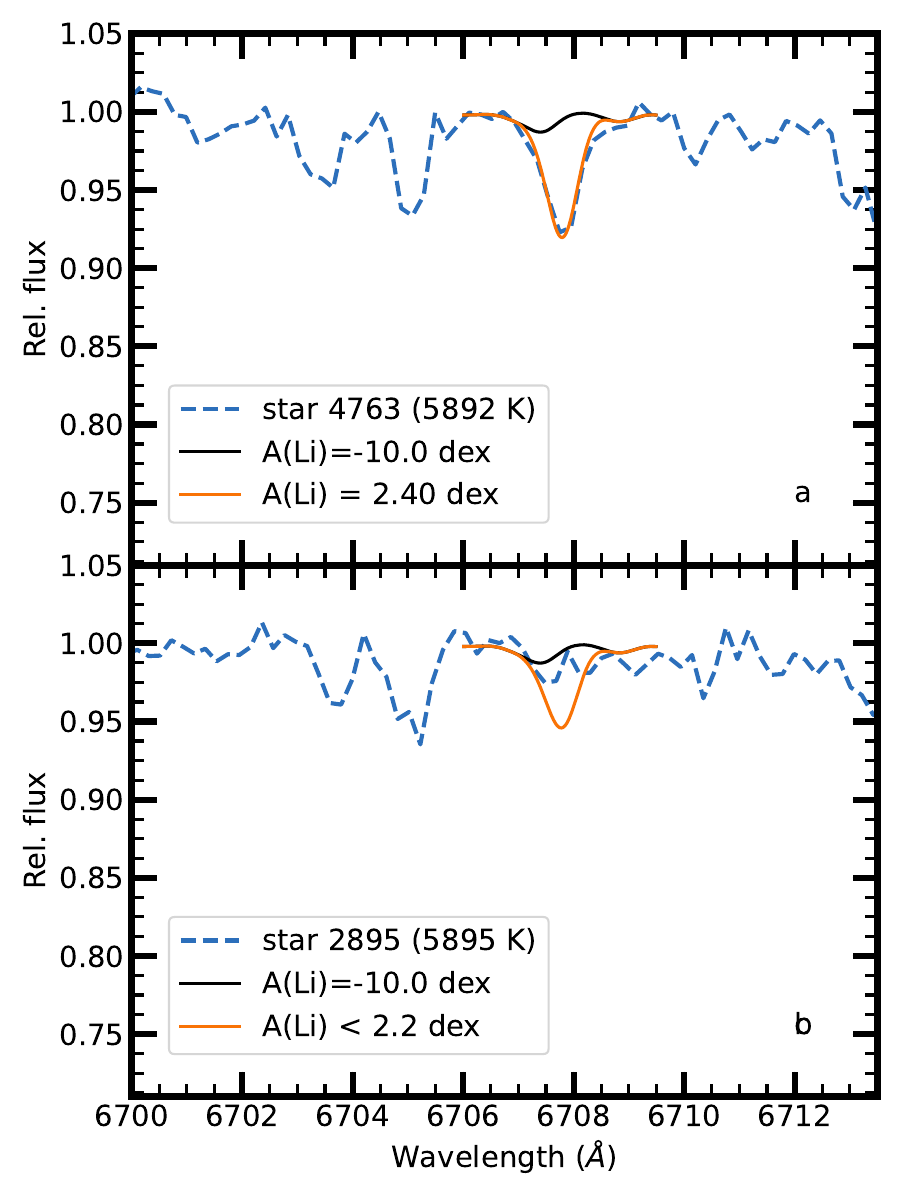}
	\caption{Spectral synthesis of two SMs near 5900 K, showing a lithium detection (star 4763) and an upper-limit case (star 2895). The blue dashed line shows the observed spectrum, the black line shows the no-Li case, and the red line shows either the best fit for the A(Li) detection or the value corresponding to the 3$\sigma$ upper limit.}
	\label{fig3}
\end{figure}

Figure~\ref{fig2} presents the observed A(Li) for MS stars in NGC 188, including both SM and BM, with either detections or upper limits. Near 5900 K, A(Li) shows significant scatter, ranging from detections around 2.6 dex down to upper limits of 2.2 dex. Shown in Figure \ref{fig3} are sample spectral syntheses for two stars near 5900 K: one with a clear detection and one with only an upper limit, demonstrating that the observed scatter is real. For comparison, we show A(Li) detections for five NGC 188 dwarfs from \citet{1988ApJ...334..734H} and eleven dwarfs from \citet{2003A&A...399..133R}. These literature values show similar patterns to our results, but share only one star in common; the remaining non-overlapping stars are mostly brighter than our magnitude cut of $V=15.4$ mag.

\section{The A(Li) -- $T_{\rm eff}$ morphology}

\begin{figure*}
	\centering
	\includegraphics[width=0.95\textwidth]{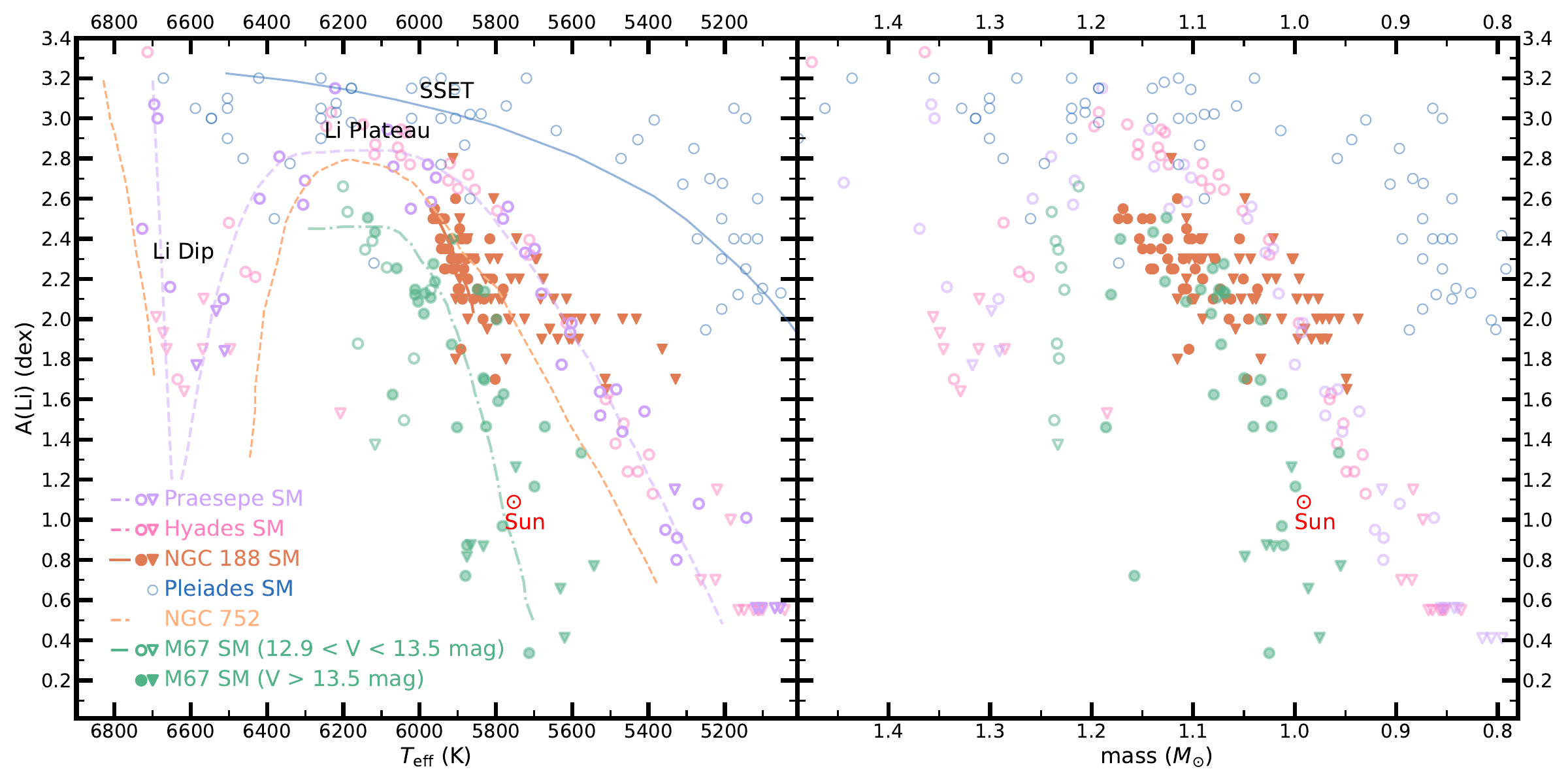}
	\caption{A(Li) as a function of $T_{\rm eff}$ (left panel) and mass (right panel) for main-sequence stars in NGC 188. The same symbols are used for NGC 188 as in Figure~\ref{fig1}, while only SM is shown. The observed data are shown as colored symbols in both panels. In the left panel, the SSET model for Pleiades is shown as a blue line. The A(Li) -- $T_{\rm eff}$ trend for Pleiades (120 Myr, \citealt{2021ApJ...908..119M, 2007PhDT.........2M, 2018A&A...613A..63B}), Hyades/Praesepe (650 Myr, \citealt{2017AJ....153..128C}), NGC 752 (1.8 Gyr, \citealt{2022ApJ...927..118B}), M67 (4 Gyr, \citealt{2020AJ....159..246S}), and NGC 188 (6.3 Gyr) are shown in colored lines as denoted in the legend. The circles are A(Li) detections, while the downward triangles are A(Li) 3$\sigma$ upper limits. For comparison, we mark the Sun at 5777 K (1 $M_{\odot}$) and A(Li) = 1.05 dex (\citealt{1997AJ....113.1871K}) using a red $\odot$ symbol. Open symbols mark upper main-sequence stars (12.9 $< V < 13.5$ mag) in M67, while filled symbols denote lower main-sequence stars ($V > 13.5$ mag).}
	\label{fig4}
\end{figure*}

For comparison, in Figure \ref{fig4} (left panel) we show A(Li) of MS stars from the younger open clusters Pleiades ($\sim$ 120 Myr, [Fe/H] = + 0.03 dex; \citealt{2021ApJ...908..119M, 2007PhDT.........2M, 2018A&A...613A..63B}), Hyades and Praesepe ($\sim$650 Myr, [Fe/H] = +0.15 dex; \citealt{2017AJ....153..128C}), the intermediate-aged cluster NGC 752 ($\sim$1.45 Gyr, [Fe/H] = -0.01 dex; \citealt{2015AJ....150..134T, 2019ApJ...878...99L, 2022ApJ...927..118B}), M67 ($\sim$4 Gyr, [Fe/H] = 0.0 dex; \citealt{2012MSAIS..22...97P}), and the Sun ($\sim$4 Gyr, [Fe/H] = 0.0 dex; \citealt{1997AJ....113.1871K, 2020AJ....159..246S}). We place all published A(Li) and $T_{\rm eff}$ values onto the same scale as NGC 188 by using the $B-V$–$T_{\rm eff}$ relation from \citet{2017AJ....153..128C}, fit the $\Delta A({\rm Li})$–$\Delta T_{\rm eff}$ relation, convert $\Delta T_{\rm eff}$ to $\Delta A({\rm Li})$, and recalibrate A(Li) to enable consistent comparison across different studies. The general A(Li)–$T_{\rm eff}$ trend for each cluster is shown using colored lines. For NGC 188, we fit this trend using only SM stars with A(Li) detections. Similarly, the trend lines for M67 and Hyades/Praesepe are based on SM detections only, and for these clusters BM are not shown. Because Hyades and Praesepe have nearly identical ages and Li patterns, we adopt a single trend line for both clusters, as supported by \citet{2017AJ....153..128C}.

We show the SSET prediction of A(Li) for Pleiades (\citealt{2015MNRAS.449.4131S}) as the blue line in Figure~\ref{fig4}. It reproduces the A(Li) observed in slow rotators from young clusters such as the Pleiades ($\sim$100 Myr) and M35 ($\sim$150 Myr), but fails to explain the Li-rich fast rotators in these clusters or the much lower A(Li) levels found in older populations. The SSET does not incorporate any additional mixing, diffusion, or mass loss processes. In the Hyades and Praesepe ($\sim$650 Myr), A(Li) spans a broad range of $T_{\rm eff}$, showing the characteristic F-dwarf Li-Dip (6320–7000 K), a relatively flat plateau between 6000–6200 K, and a sharp decrease in A(Li) among cooler G dwarfs ($T_{\rm eff} <$ 6000 K). These values are significantly lower than SSET predictions. M67 ($\sim$4 Gyr) shows a shorter and more scattered Li plateau, followed by greater depletion and more scatter in A(Li) toward cooler G dwarfs. M67 shows much lower A(Li) and a steeper A(Li)–$T_{\rm eff}$ trend for cooler G dwarfs compared to the Hyades and Praesepe. NGC 188 is even older and contains only cooler G dwarfs, no Li plateau stars remain, and all observed stars are Li depleted. The absence of Li plateau stars in NGC 188 may be due to its old age of 6.3 Gyr so that, a) its more massive stars have already evolved off of the MS, and b) the Li-$T_{\rm eff}$ relation becomes increasingly steep. 

For stars with $T_{\rm eff} < 5700$ K, we mostly report upper limits on A(Li). Among the A(Li) detections, the highest values in NGC 188 are similar to the upper envelope of Li plateau stars in M67. Cooler than the Li plateau, the least Li-depleted stars in NGC 188 lie well above the upper envelope of A(Li) seen in M67. When upper limits are considered, stars in NGC 188 at $T_{\rm eff} \sim 5900$ K span at least $\sim$0.8 dex in A(Li), which can be comparable to the scatter observed in M67. We also mark the position of the Sun in Figure~\ref{fig4}. Around the solar $T_{\rm eff}$, the mix of detections and upper limits makes it unclear whether any NGC 188 stars are as Li-poor as the Sun. Deeper observations reaching lower detection limits will be needed to assess whether the Sun is unusually Li-depleted compared to slightly older stars. NGC 188 is clearly older than M67, as its turnoff stars are less massive. This challenges the conventional expectation that older stars should show more Li depletion in a Li-$T_{\rm eff}$ diagram. In particular, some G dwarfs in NGC 188 appear only slightly more depleted in Li than those in the much younger Hyades and Praesepe clusters. The A(Li) of NGC 188 are comparable to those observed in NGC 752 (1.8 Gyrs, [Fe/H] = -0.15 dex), although the A(Li) - $T_{\rm eff}$ trend in NGC 188 is notably steeper.

In addition to $T_{\rm eff}$, we also examine the A(Li)–mass relation. Stellar masses are derived from the Yale–Yonsei isochrones (\citealt{2004ApJS..155..667D}) appropriate for the age and metallicity of each cluster, and A(Li) versus mass is shown in the right panel of Figure \ref{fig4}. In the Li–mass diagram, the overall depletion patterns of M67 and NGC 188 appear more broadly similar. The comparison is limited, however, by the small number of M67 detections near $M \approx 1.13,M_\odot$, with two lying on the upper envelope of the NGC 188 detections and one on the lower envelope. At slightly lower masses ($M \approx 1.06$–$1.09,M_\odot$), where M67 has many detections, NGC 188 provides almost exclusively upper limits, offering little constraint. At lower masses, the NGC 188 data consist only of upper limits, which are consistent with the lower envelope of M67 detections but otherwise give no information as to how the A(Li) of NGC 188 compare with those of M67. The larger scatter further complicates the comparison, and while there is a marginal suggestion that NGC 188 may lie above M67 in the $1.06$–$1.09,M_\odot$ range, the uncertainties are too large to support a firm conclusion. In general, the Li–mass diagrams are less steep than the Li–$T_{\rm eff}$ diagrams. In either representation, however, the NGC 188 stars do not lie below M67 despite the older age of the cluster, contrary to the expectation that Li depletes with age. BMs are excluded in those discussions, as they might follow a different evolutionary path compared to SMs, though our sample does not suggest a distinct difference in A(Li) between BMs and SMs.

One possible explanation is that differences in the initial conditions of the cluster environment may lead to differences in Li depletion.  In particular, differences between the two clusters in their initial angular momentum distributions could lead to differences in Li depletion.  For example, if NGC 188 formed with slightly more Li than M67 because NGC 188 is slightly more metal-rich (\citealt{2011PhDT.......192C}), and NGC 188 depleted slightly more Li than M67 because it is slightly older, then their A(Li) could be similar at a given mass (Figure 
\ref{fig4}, right panel) though at the same time NGC 188 might lie {\it above} M67 in a Li-$T_{\rm eff}$ diagram.  While age plays a key role in Li depletion, it may not be the only factor. Cluster-specific conditions and unknown aspects of stellar evolutionary histories may also contribute to the observed A(Li) patterns in open clusters such as NGC 188 and M67. 

Both theory (\citealt{1990ApJS...73...21D, 1991ApJ...370L..89D}) and observation (\citealt{1982A&A...115..357S, 1995ApJ...453..819R}) suggest that Li depletion depends on metallicity, so Li depletion cannot be explained by age alone (\citealt{2005A&A...442..615S}, \citealt{2008A&A...489..677P}, \citealt{2010IAUS..268..275R}). More recently, \citet{2023MNRAS.523..802J} analyzed a large homogeneous dataset from the Gaia-ESO Survey and showed that Li-based age estimates for G-type stars older than 1 Gyr are uncertain and cannot be explained solely by metallicity. Our results provide the most complete main-sequence Li dataset for a 6 Gyr open cluster to date, and underscore that age alone is likely insufficient to explain the observed Li patterns.

\section*{Acknowledgements}
		
Q.S. is supported by the National Key R\&D Program of China No. 2024YFA1611801, the Science and Technology Commission of Shanghai Municipality under Grant No. 25ZR1402244, and the Startup Fund for Young Faculty at Shanghai Jiao Tong University. This work has been supported by the National Science Foundation through grant AST-1909456 to C.P.D. We thank the WIYN 3.5 m staff for helping us obtain excellent spectra. This work has made use of data from the European Space Agency (ESA) mission Gaia (\url{https://www.cosmos.esa.int/gaia}), processed by the Gaia Data Processing and Analysis Consortium (DPAC, \url{https://www.cosmos.esa.int/web/gaia/dpac/consortium}). Funding for the DPAC has been provided by national institutions, in particular, the institutions participating in the Gaia Multilateral Agreement.
		
\bibliography{Li_ms}{}
\bibliographystyle{aasjournal}
		
\end{document}